\documentclass[sigconf]{acmart}

\usepackage{booktabs} 
\setcopyright{acmlicensed}

\usepackage{amsmath}
\usepackage{amssymb}
\usepackage{mathtools}
\usepackage{graphicx}
\usepackage{subcaption}
\usepackage{bm}
\usepackage{bbm}
\usepackage{url}
\usepackage{tikz}
\usetikzlibrary{bayesnet}

\DeclareMathOperator{\E}{\mathbb{E}}

\DeclarePairedDelimiterX{\infdivx}[2]{(}{)}{%
  #1\;\delimsize\|\;#2%
}
\newcommand{\infdiv}{D_{KL}\infdivx}

\begin{document}

\copyrightyear{2017} 
\acmYear{2017} 
\setcopyright{acmlicensed}
\acmConference{SIGIR '17}{}{August 07-11, 2017, Shinjuku, Tokyo, Japan}
\acmPrice{15.00}
\acmDOI{http://dx.doi.org/10.1145/3077136.3080816}
\acmISBN{978-1-4503-5022-8/17/08}

\title{Variational Deep Semantic Hashing for Text Documents}

\author{Suthee Chaidaroon}
\affiliation{Department of Computer Engineering}
\affiliation{Santa Clara University}
\affiliation{Santa Clara, CA 95053, USA}
\email{schaidaroon@scu.edu}
\author{Yi Fang}
\affiliation{Department of Computer Engineering}
\affiliation{Santa Clara University}
\affiliation{Santa Clara, CA 95053, USA}
\email{yfang@scu.edu}

\begin{abstract}
As the amount of textual data has been rapidly increasing over the past decade, efficient similarity search methods have become a crucial component of large-scale information retrieval systems. A popular strategy is to represent original data samples by compact binary codes through hashing. A spectrum of machine learning methods have been utilized, but they often lack expressiveness and flexibility in modeling to learn effective representations. The recent advances of deep learning in a wide range of applications has demonstrated its capability to learn robust and powerful feature representations for complex data. Especially, deep generative models naturally combine the expressiveness of probabilistic generative models with the high capacity of deep neural networks, which is very suitable for text modeling. However, little work has leveraged the recent progress in deep learning for text hashing. 

In this paper, we propose a series of novel deep document generative models for text hashing.  The first proposed model is unsupervised while the second one is supervised by utilizing document labels/tags for hashing. The third model further considers document-specific factors that affect the generation of words. The probabilistic generative formulation of the proposed models provides a principled framework for model extension, uncertainty estimation, simulation, and interpretability. Based on variational inference and reparameterization, the proposed models can be interpreted as encoder-decoder deep neural networks and thus they are capable of learning complex nonlinear distributed representations of the original documents. We conduct a comprehensive set of experiments on four public testbeds. The experimental results have demonstrated the effectiveness of the proposed supervised learning models for text hashing. 
\end{abstract}

%
%



\maketitle

\section{Introduction}
The task of similarity search, also known as nearest neighbor search, proximity search, or close item search, is to find similar items given a query object \cite{wang2016survey}. It has many important information retrieval applications such as document clustering, content-based retrieval, and collaborative filtering \cite{wang2016learning}. The rapid growth of Internet has resulted in massive textual data in the recent decades. In addition to the cost of storage, searching for relevant content in gigantic databases is even more daunting. Traditional text similarity computations are conducted in the original vector space and could be prohibitive to use for large-scale corpora since these methods are involved with high cost of numerical computation in the high-dimensional spaces.

Many research efforts have been devoted to approximate similarity search that is shown to be useful for practical problems. Hashing \cite{gionis1999similarity,weiss2009spectral,salakhutdinov2007semantic} is an effective solution to accelerate similarity search by designing compact binary codes in a low-dimensional space so that semantically similar documents are mapped to similar codes. This approach is much more efficient in memory and computation. A binary representation of each document often only needs 4 or 8 bytes to store, and thus a large number of encoded documents can be directly loaded into the main memory. Computing similarity between two documents can be accomplished by using bitwise XOR operation which takes only one CPU instruction. A spectrum of machine learning methods have been utilized in hashing, but they often lack expressiveness and flexibility in modeling, which prevents them from learning compact and effective representations of text documents.

On the other hand, deep learning has made tremendous progress in the past decade and has demonstrated impressive successes in a variety of domains including speech recognition, computer vision, and natural language processing \cite{lecun2015deep}. One of the main purposes of deep learning is to learn robust and powerful feature representations for complex data. Recently, deep generative models with variational inference \cite{rezende2014stochastic,kingma2014auto} have further boosted the expressiveness and flexibility for representation learning by integrating deep neural nets into the probabilistic generative framework. The seamless combination of generative modeling and deep learning makes them suitable for text hashing. However, to the best of our knowledge, no prior work has leveraged them for hashing tasks. 

In this paper, we propose a series of novel deep document generative models for text hashing, inspired by variational autoencoder (VAE) \cite{rezende2014stochastic,kingma2014auto}. The proposed models are the marriage of deep learning and probabilistic generative models \cite{blei2012probabilistic}. They enjoy the good properties of both learning paradigms. First, with the deep neural networks, the proposed models can learn flexible nonlinear distributed representations of the original high-dimensional documents. This allows individual codes to be fairly general and concise but their intersection to be much more precise. For example, nonlinear distributed representations allow the topics/codes ``government," ``mafia," and ``playboy" to combine to give very high probability to the word ``Berlusconi," which is not predicted nearly as strongly by each topic/code alone. 

Meanwhile, the proposed models are probabilistic generative models and thus there exists an underlying data generation process characterizing each model. The probabilistic generative formulation provides a principled framework for model extensions such as incorporating supervisory signals and adding private variables. The first proposed model is unsupervised and can be interpreted as a variant of variational autoencoder for text documents. The other two models are supervised by utilizing the document label/tag information. The prior work in the literature \cite{wang2013semantic} has demonstrated that the supervisory signals are crucial to boost the performance of semantic hashing for text documents. The third model further adds a private latent variable for documents to capture the information only concerned with the documents but irrelevant to the labels, which may help remove noises from document representations. Furthermore, specific constraints can be enforced by making explicit assumptions in the models. One desirable property of hash code is to ensure the bits are uncorrelated so that the next bit cannot be predicted based on the previous bits \cite{weiss2009spectral}. To achieve this property,  we can just assume that the latent variable has a prior distribution with independent dimensions. 

In sum, the probabilistic generative formulation provides a principled framework for model extensions, interpretability, uncertainty estimation, and simulation, which are often lacking in deep learning models but useful in text hashing. The main contributions of the paper can be summarized as follow:

\begin{itemize}
    \item We proposed a series of unsupervised and supervised deep document generative models to learn compact representations for text documents. To the best of our knowledge, this is the first work that utilizes deep generative models with variational inference for text hashing.
    \item The proposed models enjoy both advantages of deep learning and probabilistic generative models. They can learn complex nonlinear distributed representations of the original high-dimensional documents while providing a principled framework for probabilistic reasoning.
    \item We derived tractable variational lowerbounds for the proposed models and reparameterize the models so that backpropagation can be applied for efficient parameter estimation.
    \item We conducted a comprehensive set of experiments on four public testbeds. The experimental results demonstrate significant improvements in our supervised models over several well-known semantic hashing baselines. 
\end{itemize}

\section{Related Work}

\subsection{Hashing}

Due to computational and storage efficiencies of compact binary codes, hashing methods have been widely used for similarity search, which is an essential component in a variety of large-scale information retrieval systems \cite{wang2016learning,wang2016survey}. Locality-Sensitive Hashing (LSH) \cite{datar2004locality} is one of the most popular hashing methods with interesting asymptotic theoretical properties leading to performance guarantees. While LSH is a data-independent hashing method, many hashing methods have been recently proposed to leverage machine learning techniques with the goal of learning data-dependent hash functions, ranging from unsupervised and supervised to semi-supervised settings. Unsupervised hashing methods attempt to integrate the data properties, such as distributions and manifold structures to design compact hash codes with improved accuracy. For instance, Spectral Hashing (SpH) \cite{weiss2009spectral} explores the data distribution by preserving the similarity between documents by forcing the balanced and uncorrelated constraints into the learned codes, which can be viewed as an extension of spectral clustering \cite{ng2001spectral}. Graph hashing \cite{liu2011hashing} utilizes the underlying manifold structure of data captured by a graph representation. Self Taught Hashing (STH) \cite{zhang2010self} is the state-of-the-art hashing method by decomposing the learning procedure into two steps: generating binary code and learning hash function. 

Supervised hashing methods attempt to leverage label/tag information for hash function learning. It has attracted more and more attention in recent years. For example, Wang et al. \cite{wang2013semantic} propose Semantic Hashing using Tags and Topic Modeling (SHTTM) to incorporate tags to obtain more effective hashing codes via a matrix factorization formulation. To utilize the pairwise supervision information in the hash function learning, Kernel-Based Supervised Hashing (KSH) proposed in \cite{liu2012supervised} used a pairwise relationship between samples to achieve high-quality hashing. Binary Reconstructive Embedding (BRE) \cite{kulis2009learning} was proposed to learn hash functions by minimizing the reconstructed error between the metric space and Hamming space. Moreover, there are also several works using the ranking order information to design hash functions. Ranking-based Supervised Hashing (RSH) \cite{wang2013learning} was proposed to leverage listwise supervision into the hash function learning framework. Semi-supervised learning paradigm was also employed to design hash functions by using both labeled and unlabeled data \cite{wang2010semi}. The hashing-code learning problem is essentially a discrete optimization problem which is difficult to solve. Most existing supervised hashing methods try to solve a relaxed continuous optimization problem and then threshold the continuous representation to obtain a binary code. Abundant related work, especially on image hashing, exists in the literature. Two recent surveys \cite{wang2016learning,wang2016survey} provide a comprehensive literature review.





\subsection{Deep Learning}

Deep learning has drawn increasing attention and research efforts in a variety of artificial intelligence areas including speech recognition, computer vision, and natural language processing. Since one main purpose of deep learning is to learn robust and powerful feature representations for complex data, it is very natural to leverage deep learning for exploring compact hash codes which can be regarded as binary representations of data. Most of the related work has focused on image data \cite{xia2014supervised, lai2015simultaneous, erin2015deep, lin2015deep} rather than text documents probably due to the effectiveness of the convolution neural networks (CNNs) to learn good low-dimensional representations of images. The typical deep learning architectures for hash function learning consist of CNNs layers for representation learning and hash function layers which then transform the representation to supervisory signals. The loss functions could be pointwise \cite{lin2015deep}, pairwise \cite{erin2015deep}, or listwise \cite{lai2015simultaneous}. 


Some recent works have applied deep learning for several IR tasks such as ad-hoc retrieval \cite{guo2016deep}, web search \cite{huang2013learning}, and ranking pairs of short texts \cite{severyn2015learning}. However, very few has investigated deep learning for text hashing. The representative work is semantic hashing \cite{salakhutdinov2007semantic}. It builds a stack of restricted Boltzmann machines (RBMs) \cite{hinton2012practical} to discover hidden binary units which can model input text data (i.e., word-count vectors). After learning a multilayer RBM through pretraining and fine tuning on a collection of documents, the hash code of any document is acquired by simply thresholding the output of the deepest layer. A recent work \cite{xu2015convolutional} exploited convolutional neural network for text hashing, which relies on external features such as the GloVe word embeddings to construct text representations. 

Recently, deep generative models have made impressive progress with the introduction of the variational autoencoders (VAEs) \cite{rezende2014stochastic,kingma2014auto} and Generative Adversarial Networks (GANs) \cite{goodfellow2014generative}. VAEs are especially an appealing framework for generative modeling by coupling the approach of variational inference \cite{wainwright2008graphical} with deep learning. As a result, they enjoy the advantages of both deep learning and probabilistic graphical models. Deep generative models parameterized by neural networks have achieved state-of-the-art performance in unsupervised and supervised learning \cite{kingma2014auto,kingma2014semi,miao2015neural}. To the best of our knowledge, our proposed models are the first work that utilizes variational inference with deep learning for text hashing. It is worth pointing out that both semantic hashing with stacked RBMs \cite{salakhutdinov2007semantic} and our models are deep generative models, but the former is undirected graphical models, and the latter is directed models. The underlying generative process of directed probabilistic models makes them easy to interpret and extend. The proposed models are very scalable since they are trained as deep neural networks by efficient backpropagation, while the stacked RBMs are often much harder to train \cite{hinton2012practical}.

\begin{figure*}[t]
\centering
    \includegraphics[width=18cm]{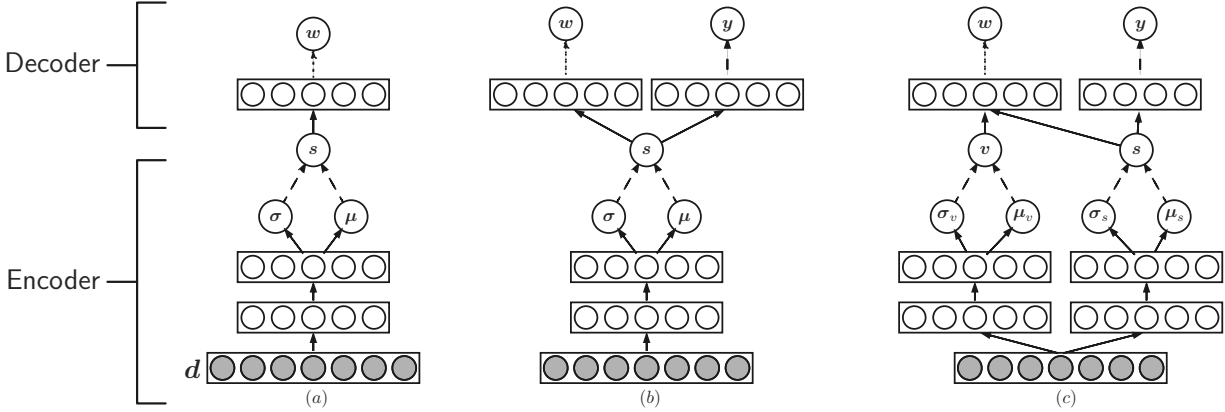}  
    \caption{Architectures of (a) VDSH, (b) VDSH-S, and (c) VDSH-SP. The dashed line represents a stochastic layer.}\label{architecture}
    \label{fig:arch}
\end{figure*}

\section{Variational Deep Semantic Hashing}

This section presents three novel deep document generative models to learn low-dimensional semantic representations of documents for text hashing. In Section \ref{model-VDSH}, we introduce the basic model which is essentially a variational autoencoder for text modeling. Section \ref{model-VDSH-S} extends the model to utilize label information to learn a more sensible representation. Section \ref{model-VDSH-SP} further incorporates document private variables to model document-specific information. Based on the variational inference, all the three models can be viewed as having an encoder-decoder neural network architecture where the encoder compresses a high-dimensional document to a compact latent semantic vector and the decoder reconstructs the document (or the labels). Section \ref{binhcode} discusses two thresholding methods to convert the continues latent vector to a binary code for text hashing.

\subsection{Unsupervised Learning (VDSH)}
\label{model-VDSH}

In this section, we present the basic variational deep semantic hashing (VDSH) model for the unsupervised learning setting. VDSH is a probabilistic generative model of text which aims to extract a continuous low-dimensional semantic representation $\boldsymbol{s} \in \mathbb{R}^K$ for each document. Let $\boldsymbol{d} \in \mathbb{R}^V$ be the bag-of-words representation of a document and $\boldsymbol{w}_i \in \{0, 1\}^V$ be the one-hot vector representation of the $i^{th}$ word of the document where $V$ is the vocabulary size. $\boldsymbol{d}$ could be represented by different term weighting schemes such as binary, TF, and TFIDF \cite{manning2008introduction}. The document generative process can be described as follows:



\begin{itemize}
\item For each document $\boldsymbol{d}$,
\begin{itemize}
		\item Draw a latent semantic vector $\boldsymbol{s} \sim P(\boldsymbol{s})$ where $P(\boldsymbol{s}) = \mathcal{N}(\boldsymbol{0}, \boldsymbol{I})$ is the standard Gaussian distribution.
		\item For the $i^{th}$ word in the document, 
		\begin{itemize}
		\item Draw $\boldsymbol{w}_i \sim P(\boldsymbol{w}_i | f(\boldsymbol{s}; \theta))$.	
		\end{itemize}
\end{itemize}
\end{itemize}
The conditional probability over words $\boldsymbol{w}_i$ is modelled by multinomial logistic regression and shared across documents as below:
\begin{eqnarray}
P(\boldsymbol{w}_i | f(\boldsymbol{s}; \theta)) = \frac{\exp(\boldsymbol{w}_i^Tf(\boldsymbol{s}; \theta))}{\sum_{j=1}^V \exp(\boldsymbol{w}_j^Tf(\boldsymbol{s}; \theta))} 
\end{eqnarray}

While $P(\boldsymbol{s})$ is a simple Gaussian distribution, any distribution can
be generated by mapping the simple Gaussian through a sufficiently complicated function \cite{doersch2016tutorial}. Thus, $f(\boldsymbol{s}; \theta))$ is such a highly flexible function approximator usually a neural network. In other words, we can learn a function which maps our independent, normally-distributed $\boldsymbol{s}$ values to whatever latent semantic variables might be needed for the model, and then generate the word $\boldsymbol{w}_i$. However, introducing a highly nonlinear mapping from $\boldsymbol{s}$ to $\boldsymbol{w}_i$ results in intractable data likelihood $ \int_{\boldsymbol{s}} P(\boldsymbol{d}| \boldsymbol{s}) P(\boldsymbol{s})d\boldsymbol{s}$ and thus intractable posterior distribution $P(\boldsymbol{s}|\boldsymbol{d})$ \cite{kingma2014auto}. Similar to VAE, we use an approximation $Q(\boldsymbol{s}|\boldsymbol{d}; \phi)$ for the true posterior distribution. By applying the variational inference principle \cite{wainwright2008graphical}, we can obtain the following tractable lowerbound of the document log likelihood (see \cite{kingma2014auto} and Appendix):

\begin{equation}
\mathcal{L}_1 = \E_Q \big[\sum_{i=1}^N \log P(\boldsymbol{w}_i | f(\boldsymbol{s}; \theta))\big] - \infdiv{Q(\boldsymbol{s}|\boldsymbol{d}; \phi)}{P(\boldsymbol{s})}\label{lowerbound1}
\end{equation}
where $N$ is the number of words in the document and $\infdiv{}{}$ is the Kullback-Leibler (KL) divergence between the approximate posterior distribution $Q(\boldsymbol{s}|\boldsymbol{d}; \phi)$ and the prior $P(\boldsymbol{s})$. The variational distribution $Q(\boldsymbol{s}|\boldsymbol{d}; \phi)$ acts as a proxy to the true posterior $P(\boldsymbol{s}|\boldsymbol{d})$. To enable a high capacity, it is assumed to be a Gaussian $\mathcal{N}(\boldsymbol{\mu}, \text{diag}(\boldsymbol{\sigma^2}))$ whose mean $\boldsymbol{\mu}$ and variance $\boldsymbol{\sigma^2}$ are the output of a highly nonlinear function of $\boldsymbol{d}$ denoted as $g(\boldsymbol{d}; \phi)$ parameterized by $\phi$, once again typically a neural network. 

In training, the variational lowerbound in Eqn.(\ref{lowerbound1}) is maximized with respect to the model parameters. Since $P(\boldsymbol{s})$ is a standard Gaussian prior, the KL Divergence $\infdiv{Q(\boldsymbol{s}|\boldsymbol{d}; \phi)}{P(\boldsymbol{s})}$ in Eqn.(\ref{lowerbound1}) can be computed analytically. The first term $\E_Q$ can be viewed as an expected negative reconstruction error of the words in the document and it can be computed based on the Monte Carlo estimation \cite{goodfellow2016deep}.

Based on Eqn.(\ref{lowerbound1}), we can interpret VDSH as a variational autoencoder with discrete output: a feedforward neural network encoder $Q(\boldsymbol{s}|\boldsymbol{d}; \phi)$ compresses document representations into continuous hidden vectors, i.e., $\boldsymbol{d} \rightarrow \boldsymbol{s}$; a softmax decoder $\sum_{i=1}^N P(\boldsymbol{w}_i | f(\boldsymbol{s}; \theta))$ reconstructs the documents by independently generating the words $\boldsymbol{s} \rightarrow \{\boldsymbol{w}_i\}_{i=1}^N$. Figure \ref{architecture}(a) illustrates the architecture of VDSH. In the experiments, we use the following specific architecture for the encoder and decoder.
\begin{equation*}
\begin{aligned}[c]
\text{Encoder} \quad Q(\boldsymbol{s}|g(\boldsymbol{d}; \phi)):\\
\bm{t}_1 = \text{ReLU}(\bm{W}_1\bm{d}+\bm{b}_1)\\
\bm{t}_2 = \text{ReLU}(\bm{W}_2\bm{t}_1+\bm{b}_2)\\
\bm{\mu} = \bm{W}_3\bm{t}_2 + \bm{b}_3\\
\log \bm{\sigma} = \bm{W}_4\bm{t}_2+\bm{b}_4\\
\bm{s} \sim \mathcal{N}(\bm{\mu(d)}, \text{diag}(\bm{\sigma^2(d))})
\end{aligned}
\qquad
\begin{aligned}[c]
\text{Decoder} \quad P(\boldsymbol{w}_i | f(\boldsymbol{s}; \theta)):\\
\bm{c}_i = \exp(-\bm{s}^T \bm{G} \bm{w}_i + \bm{b}_{\bm{w}_i})\\
P(\bm{w}_i|\bm{s}) = \frac{\bm{c}_i}{\sum_{k=1}^V \bm{c}_k}\\
P(\bm{d}|\bm{s}) = \prod_{i=1}^N P(\bm{w}_i|\bm{s})
\end{aligned}
\end{equation*}
This architecture is similar to the one presented in VAE \cite{rezende2014stochastic} except that VDSH has the softmax layer to model discrete words while VAE is proposed to model images as continuous output. Here, the encoder has two Rectified Linear Unit (ReLU) \cite{goodfellow2016deep} layers. ReLU generally does not face gradient vanishing problem as with other activation functions. Also, it has been shown that deep neural networks can be trained efficiently using ReLU even without pretraining \cite{goodfellow2016deep}. 

In this architecture, there is a stochastic layer which is to sample $\boldsymbol{s}$ from a Gaussian distribution $\mathcal{N}(\bm{\mu(d)}, \text{diag}(\bm{\sigma^2(d))})$, as represented by the dashed lines in the middle of the networks in Figure \ref{architecture}. Backpropagation cannot handle stochastic layer within the network. In practice, we can leverage the ``location-scale" property of Gaussian distribution, and use the reparameterization trick \cite{kingma2014auto} to turn the stochastic layer of $\boldsymbol{s}$ to be deterministic. As a result, the encoder $Q(\boldsymbol{s}|\boldsymbol{d}; \phi)$ and decoder $P(\boldsymbol{w}_i | f(\boldsymbol{s}; \theta))$ form an end-to-end neural network and are then trained jointly by maximizing the variational lowerbound in Eqn.(\ref{lowerbound1}) with respect to their parameters by the standard backpropagation algorithm \cite{goodfellow2016deep}. 




\subsection{Supervised Learning (VDSH-S)}
\label{model-VDSH-S}

In many real-world applications, documents are often associated with labels or tags which may provide useful guidance in learning effective hashing codes. Document content similarity in the
original bag-of-word space may not fully reflect the semantic relationship between documents. For example, two documents in the same category may have low document content similarity due to the vocabulary gap, while their semantic similarity could be high. In this section, we extend VDSH to the supervised setting with the new model denoted as VDSH-S. The probabilistic generative process of a document with labels is as follows:

\begin{itemize}
\item For each document $\boldsymbol{d}$,
\begin{itemize}
		\item Draw a latent semantic vector $\boldsymbol{s} \sim P(\boldsymbol{s})$ where $P(\boldsymbol{s}) = \mathcal{N}(\boldsymbol{0}, \boldsymbol{I})$ is the standard Gaussian distribution.
		\item For the $i^{th}$ word in the document, 
		\begin{itemize}
		\item Draw $\boldsymbol{w}_i \sim P(\boldsymbol{w}_i | f(\boldsymbol{s}; \theta))$.	
		\end{itemize}
		\item For the $j^{th}$ label in the label set, 
		\begin{itemize}
		\item Draw $\boldsymbol{y}_j \sim P(\boldsymbol{y} | f(\boldsymbol{s}; \tau))$.
		\end{itemize}
\end{itemize}
\end{itemize}
where $\boldsymbol{y}_j \in \{0, 1\}^L$ is the one-hot representation of the label $j$ in the label set and $L$ is the total number of possible labels (the size of the label set). Let us use $\boldsymbol{Y} \in \{0, 1\}^L$ represent the bag-of-labels of the document (i.e., if the document has label $j$, the $j^{th}$ dimension of $\boldsymbol{Y}$ is 1; otherwise, it is 0). VDSH-S assumes that both words and labels are generated based on the same latent semantic vecotor.

We assume a general multi-label classification setting where each document could have multiple labels/tags.  $P(\boldsymbol{y}_j | f(\boldsymbol{s}; \tau))$ can be modeled by the logistic function as follows:
\begin{equation}
P(\boldsymbol{y}_j | f(\boldsymbol{s}; \tau)) = \frac{1}{1+\exp(-\boldsymbol{y}_j^T f(\boldsymbol{s}; \tau))}
\end{equation}
Similar to VDSH, $f(\boldsymbol{s}; \tau)$ is parameterized by a neural network with the parameter $\tau$ so that we can learn an effective mapping from the latent semantic vector to the labels. The lowerbound of the data log likelihood can be similarly derived and shown as follows:
\begin{eqnarray}
\mathcal{L}_2 = \E_Q \big[ \sum_{i=1}^N \log P(\boldsymbol{w}_i | f(\boldsymbol{s}; \theta)) + \sum_{j=1}^L \log P(\boldsymbol{y}_j | f(\boldsymbol{s}; \tau))\big] \nonumber\\
- \infdiv{Q(\boldsymbol{s}|\boldsymbol{d}, \boldsymbol{Y}; \phi)}{P(\boldsymbol{s})}\label{lowerbound2}
\end{eqnarray}
Compared to Eqn.(\ref{lowerbound1}) in VDSH, this lowerbound has an extra term, $\E_Q \big[ \sum_{j=1}^L \log P(\boldsymbol{y}_j | f(\boldsymbol{s}; \tau))\big]$, which can be computed in a similar way with $\E_Q \big[\sum_{i=1}^N \log P(\boldsymbol{w}_i | f(\boldsymbol{s}; \theta))\big]$ in Eqn.(\ref{lowerbound1}), by using the Monte Carlo estimation. In addition, we can drop the dependence on variable $\boldsymbol{Y}$ in the variational distribution $Q(\boldsymbol{s}|\boldsymbol{d}, \boldsymbol{Y}; \phi)$ since we may not have the label information available for new documents.

The architecture of the VDSH-S model is shown in Figure \ref{architecture}(b). It consists of a feedforward neural network encoder of a document $\boldsymbol{d} \rightarrow \boldsymbol{s}$ and a decoder of the words and labels of the document $\boldsymbol{s} \rightarrow \{\boldsymbol{w}_i\}_{i=1}^N; \{\boldsymbol{y}_j\}_{j=1}^L$. It is worth pointing out that the labels still affect the learning of latent semantic vector by their presence in the decoder despite their absence in the encoder. By using the reparameterization trick, the model becomes a deterministic deep neural network and the lowerbound in Eqn.(\ref{lowerbound2}) can be maximized by backpropagation (see Appendix).

\subsection{Document-specific Modeling (VDSH-SP)}
\label{model-VDSH-SP}

VDSH-S assumes both document and labels are generated by the same latent semantic vector $\boldsymbol{s}$. In some cases, this assumption may be restrictive. For example, the original document may contain information that is irrelevant to the labels. It could be difficult to find a common representation for both documents and labels. This observation motivates us to introduce a document private variable $\bm{v}$, which is not shared by the labels $\bm{Y}$. The generative process is described as follows:

\begin{itemize}
\item For each document $\boldsymbol{d}$,
\begin{itemize}
		\item Draw a latent semantic vector $\boldsymbol{s} \sim P(\boldsymbol{s})$ where $P(\boldsymbol{s}) = \mathcal{N}(\boldsymbol{0}, \boldsymbol{I})$ is the standard Gaussian distribution.
		\item Draw a latent private vector $\boldsymbol{v} \sim P(\boldsymbol{v})$ where $P(\boldsymbol{v}) = \mathcal{N}(\boldsymbol{0}, \boldsymbol{I})$ is the standard Gaussian distribution.
		\item For the $i^{th}$ word in the document, 
		\begin{itemize}
		\item Draw $\boldsymbol{w}_i \sim P(\boldsymbol{w}_i | f(\boldsymbol{s}+\boldsymbol{v}; \theta))$.	
		\end{itemize}
		\item For the $j^{th}$ label in the label set, 
		\begin{itemize}
		\item Draw $\boldsymbol{y}_j \sim P(\boldsymbol{y} | f(\boldsymbol{s}; \tau))$.
		\end{itemize}
\end{itemize}
\end{itemize}
As we can see, $\bm{s}$ models the shared information between document and labels while $\bm{v}$ only contains the document-specific information. We can view adding private variables as removing the noise from the original content that is irrelevant to the labels. With the added private variable, we denote this model as VDSH-SP. The tractable variational lowerbound of data likelihood can be derived as follows:


\begin{eqnarray}
\mathcal{L}_3 = \E_Q \big[ \sum_{i=1}^N \log P(\boldsymbol{w}_i | f(\boldsymbol{s}+\boldsymbol{v}; \theta)) + \sum_{j=1}^L \log P(\boldsymbol{y}_j | f(\boldsymbol{s}; \tau))\big] \nonumber\\
- \infdiv{Q(\boldsymbol{s}|\boldsymbol{d}; \phi)}{P(\boldsymbol{s})} - \infdiv{Q(\boldsymbol{v}|\boldsymbol{d}; \phi)}{P(\boldsymbol{v})}\label{lowerbound3}
\end{eqnarray}

Similar to the other two models, VDSH-SP can be viewed as a deep neural network by applying variational inference and reparametrization. The architecture is shown in Figure \ref{architecture}(c). The Appendix contains the detailed derivations of the model.

\subsection{Binary Hash Code}
\label{binhcode}
Once a VDSH model has been trained, we can generate a compact continuous representation for any new document $\bm{d}_{new}$ by the encoder function $\mu_{new}=g(\bm{d}_{new}; \phi)$, which is the mean of the distribution $Q(\boldsymbol{s}|\boldsymbol{d}; \phi)$. The binary hashing code can then be obtained by thresholding $\mu_{new}$. The most common method of thresholding for binary code is to take the median value of the latent semantic vector $\mu$ in the training data \cite{wang2013semantic}. The rationale is based on the maximum entropy principle for efficiency which yields balanced partitioning of the whole dataset \cite{weiss2009spectral}. Thus, we set the threshold for binarizing the $p^{th}$ bit to be the median of the $p^{th}$ dimension of $\bm{s}$ in the training data. If the $p^{th}$ bit of document latent semantic vector $\mu_{new}$ is larger than the median, the $p^{th}$ binary code is set to 1; otherwise, it is set to -1. Another popular thresholding method is to use the Sign function on $\mu_{new}$, i.e., if the $p^{th}$ bit of $\mu_{new}$ is nonnegative, the corresponding code is 1; otherwise, it is -1. Since the prior distribution of the latent semantic vector is zero mean, the Sign function is also a reasonable choice. We use the median thresholding as the default method in our experiments, while also investigate the Sign function in Section \ref{thresholding}.




\subsection{Discussions}
The computational complexity of VDSH for a training document is $O(BD^2 + DSV)$. Here, $O(BK^2)$ is the cost of the encoder, where $B$ is the number of the layers in the encoder network and $D$ is the average dimension of these layers. $O(DNV)$ is the cost of the decoder, where $S$ is the
average length of the documents and $V$ is the vocabulary size. The computational complexity of VDSH-S and VDSH-SP is $O(BD^2 + DS(V+L))$ where $L$ is the size of the label set. The computational cost of the proposed models is at the same level as the deterministic autoencoder. Model learning could be quite efficient since the computations of all the models can be parallelized in GPUs, and only one sample is required during the training process. 

The proposed deep generative model has a few desirable properties for text hashing. First of all, it has the capacity of deep neural networks to learn sophisticated semantic representations for text documents. Moreover, being generative models brings huge advantages over other deep learning models such as Convolutional Neural Network (CNN) because the underlying document generative process makes the model assumptions explicit. For example, as shown in \cite{weiss2009spectral}, it is desirable to have independent feature dimensions in hash codes. To achieve this, our models just need to assume the latent semantic vector is drawn from a prior distribution with independent dimensions (e.g., standard Gaussian). The probabilistic approach also provides a principled framework for model extensions as evident in VDSH-S and VDSH-SP. Furthermore, instead of learning a particular latent semantic vector, our models learn probability distributions of the semantic vector. This can be viewed as finding a region instead of a fixed point in the latent space for document representation, which leads to more robust models. Compared with other deep generative models such as stacked RBMs and GANs, our models are computationally tractable and stable and can be estimated by the efficient backpropagation algorithm. 







\begin{table*}[ht]
\centering
\begin{tabular}{c|ccccc|ccccc}
\hline
& \multicolumn{5}{c|}{\textit{RCV1}} & \multicolumn{5}{c}{\textit{Reuters}} \\ 
\hline
Methods & 8 bits & 16 bits & 32 bits & 64 bits & 128 bits & 8 bits & 16 bits & 32 bits & 64 bits & 128 bits \\   
\hline
LSH \cite{datar2004locality} & 0.4180 & 0.4352 & 0.4716 & 0.5214 & 0.5877 & 0.2802 & 0.3215 & 0.3862 & 0.4667 & 0.5194   \\
SpH \cite{weiss2009spectral} & 0.5093 & 0.7121 & 0.7475 & 0.7559 & 0.7423 & 0.6080 & 0.6340 & 0.6513 & 0.6290 & 0.6045 \\ 
STHs \cite{zhang2010self} & 0.3975 & 0.4898 & 0.5592 & 0.5945 & 0.5946 & 0.6616 & 0.7351 & 0.7554 & 0.7350 & 0.6986  \\ 
Stacked RBMs \cite{salakhutdinov2007semantic} & 0.5106 & 0.5743 & 0.6130 & 0.6463 & 0.6531 & 0.5113 & 0.5740 & 0.6154 & 0.6177 & 0.6452 \\ 
KSH \cite{liu2012supervised} & 0.9126 & 0.9146 & 0.9221 & 0.9333 & 0.9350 & 0.7840 & 0.8376 & 0.8480 & 0.8537 & 0.8620 \\ 
SHTTM \cite{wang2013semantic} & 0.8820 & 0.9038 & 0.9258 & 0.9459 & 0.9447 & 0.7992 & 0.8520 & 0.8323 & 0.8271 & 0.8150 \\
\textbf{VDSH} & 0.7976 & 0.7944 & 0.8481 & 0.8951 & 0.8444 & 0.6859 & 0.7165 & 0.7753 & 0.7456 & 0.7318 \\
\textbf{VDSH-S} & 0.9652$\dagger$ & 0.9749$\dagger$ & \textbf{0.9801}$\dagger$ & 0.9804$\dagger$ & \textbf{0.9800}$\dagger$ & \textbf{0.9005}$\dagger$ & 0.9121$\dagger$ & \textbf{0.9337}$\dagger$ & \textbf{0.9407}$\dagger$ & 0.9299$\dagger$ \\
\textbf{VDSH-SP} & \textbf{0.9666}$\dagger$ & \textbf{0.9757}$\dagger$ & 0.9788$\dagger$ & \textbf{0.9805}$\dagger$ & 0.9794$\dagger$ & 0.8890$\dagger$ & \textbf{0.9326}$\dagger$ & 0.9283$\dagger$ & 0.9286$\dagger$ & \textbf{0.9395}$\dagger$  \\
\hline
& \multicolumn{5}{c|}{\textit{20Newsgroups}} & \multicolumn{5}{c}{\textit{TMC}} \\
\hline
Methods & 8 bits & 16 bits & 32 bits & 64 bits & 128 bits & 8 bits & 16 bits & 32 bits & 64 bits & 128 bits \\         
\hline
LSH \cite{datar2004locality} & 0.0578 & 0.0597 & 0.0666 & 0.0770 & 0.0949 & 0.4388 & 0.4393 & 0.4514 & 0.4553 & 0.4773  \\
SpH \cite{weiss2009spectral} & 0.2545 & 0.3200 & 0.3709 & 0.3196 & 0.2716 & 0.5807 & 0.6055 & 0.6281 & 0.6143 & 0.5891\\ 
STH \cite{zhang2010self} & 0.3664 & 0.5237 & 0.5860 & 0.5806 & 0.5443 & 0.3723 & 0.3947 & 0.4105 & 0.4181 & 0.4123\\ 
Stacked RBMs \cite{salakhutdinov2007semantic} & 0.0594 & 0.0604 & 0.0533 & 0.0623 & 0.0642 & 0.4846 & 0.5108 & 0.5166 & 0.5190 & 0.5137\\ 
KSH \cite{liu2012supervised} & 0.4257 & 0.5559 & 0.6103 & 0.6488 & 0.6638 & 0.6608 & 0.6842 & 0.7047 & 0.7175 & 0.7243 \\
SHTTM \cite{wang2013semantic} & 0.2690 & 0.3235 & 0.2357 & 0.1411 & 0.1299 & 0.6299 & 0.6571 & 0.6485 & 0.6893 & 0.6474 \\
\textbf{VDSH} & 0.3643 & 0.3904 & 0.4327 & 0.1731 & 0.0522 & 0.4330 & 0.6853 & 0.7108 & 0.4410 & 0.5847 \\
\textbf{VDSH-S} & 0.6586$\dagger$ & \textbf{0.6791}$\dagger$ & \textbf{0.7564}$\dagger$ & 0.6850$\dagger$ & 0.6916$\dagger$ & 0.7387$\dagger$ & \textbf{0.7887}$\dagger$ & 0.7883$\dagger$ & \textbf{0.7967}$\dagger$ & \textbf{0.8018}$\dagger$ \\
\textbf{VDSH-SP} & \textbf{0.6609}$\dagger$ & 0.6551$\dagger$ & 0.7125$\dagger$ & \textbf{0.7045}$\dagger$ & \textbf{0.7117}$\dagger$ & \textbf{0.7498}$\dagger$ & 0.7798$\dagger$ & \textbf{0.7891}$\dagger$ & 0.7888$\dagger$ & 0.7970$\dagger$ \\
\hline
\end{tabular}
\caption{Precision of the top 100 retrieved documents on four datasets with different numbers of hashing bits. The bold font denotes the best result at that number of bits. $\dagger$ denotes the improvement over the best result of the baselines is statistically significant based on the paired t-test (p-value $<0.01$).}
\label{table:main_experiments}
\end{table*} 

\section{Experimental Setup}

\subsection{Data Collections}
We use the following four public document collections for evaluation. 1) \textit{Reuters Corpus Volume I (RCV1)}. It is a large collection of manually labeled 800,000 newswire stories provided by Reuters. There are totally 103 classes. We use the full-topics version available at the LIBSVM website\footnote{\url{https://www.csie.ntu.edu.tw/~cjlin/libsvmtools/datasets/multilabel.html}}. 2) \textit{Reuters21578}\footnote{\url{http://www.nltk.org/book/ch02.html}}. A widely used text corpus for text classification. This collection has 10,788 documents with 90 categories and 7,164 unique words. 3) \textit{20Newsgroups}\footnote{\url{http://ana.cachopo.org/datasets-for-single-label-text-categorization}}. This dataset is a collection of 18,828 newsgroup posts, partitioned (nearly) evenly across 20 different newsgroups/categories. It has become a popular dataset for experiments in text applications of machine learning techniques. 4) \textit{TMC}\footnote{\url{https://catalog.data.gov/dataset/siam-2007-text-mining-competition-dataset}}. This dataset contains the air traffic reports provided by NASA and was used as part of the SIAM text mining competition. It has 22 labels, 21,519 training documents, 3,498 test documents, and 3,498 documents for the validation set. All the datasets are multi-label except \textit{20Newsgroups}. 

Each dataset was split into three subsets with roughly 80\% for training, 10\% for validation, and 10\% for test. The training data is used to learn the mapping from document to hash code. Each document in the test set is used to retrieve similar documents based on the mapping, and the results are evaluated. The validation set is used to choose the hyperparameters. We removed the stopwords using SMART's list of 571 stopwords\footnote{http://www.lextek.com/manuals/onix/stopwords2.html}. No stemming was performed. We use TFIDF \cite{manning2008introduction} as the default term weighting scheme for the raw document representation (i.e., $\boldsymbol{d}$). We experiment with other term weighting schemes in Section \ref{effectTFIDF}.

\subsection{Baselines and Settings}

We compare the proposed models with the following six competitive baselines which have been extensively used for text hashing in the prior work \cite{wang2013semantic}: Locality Sensitive Hashing (LSH)\footnote{\url{http://pixelogik.github.io/NearPy/}} \cite{datar2004locality}, Spectral Hashing (SpH)\footnote{\url{http://www.cs.huji.ac.il/~yweiss/SpectralHashing/}} \cite{weiss2009spectral}, Self-taught Hashing (STH)\footnote{\url{http://www.dcs.bbk.ac.uk/~dell/publications/dellzhang_sigir2010/sth_v1.zip}} \cite{zhang2010self}, Stacked Restricted Boltzmann Machines (Stacked RBMs) \cite{salakhutdinov2007semantic}, Supervised Hashing with Kernels (KSH) \cite{liu2012supervised}, and Semantic Hashing using Tags and Topic Modeling (SHTTM) \cite{wang2013semantic}. We used the validation dataset to choose the hyperparameters for the baselines.

For our proposed models, we adopt the method in \cite{glorot2010understanding} for weight initialization. The Adam optimizer \cite{kingma2014adam} with the step size 0.001 is used due to its fast convergence. Following the practice in \cite{wang2016deep}, we use the dropout technique \cite{srivastava2014dropout} with the keep probability of 0.8 in training to alleviate overfitting. The number of hidden nodes of the models is 1,500 for RCV1 and 1,000 for the other three smaller datasets.  All the experiments were conducted on a server with 2 Intel E5-2630 CPUs and 4 GeForce GTX TITAN X GPUs. The proposed deep models were implemented on the Tensorflow\footnote{https://www.tensorflow.org/} platform. For the VDSH model on the Reuters21578, 20Newsgroups, and TMC datasets, each epoch takes about 60 seconds, and each run takes 30 epochs to converge. For RCV1, it takes about 3,600 seconds per epoch and needs fewer epochs (about 15) to get satisfactory performance. Since RCV1 is much larger than the other three datasets, this shows that the proposed models are quite scalable. VDSH-S and VDSH-SP take slightly more time to train than VDSH does (about 40 minutes each on Reuters21578, 20Newsgroups, and TMC, and 20 hours on RCV1). 

\begin{figure*}[t]    
\hspace*{-0.5cm}
    \includegraphics[scale=0.7]{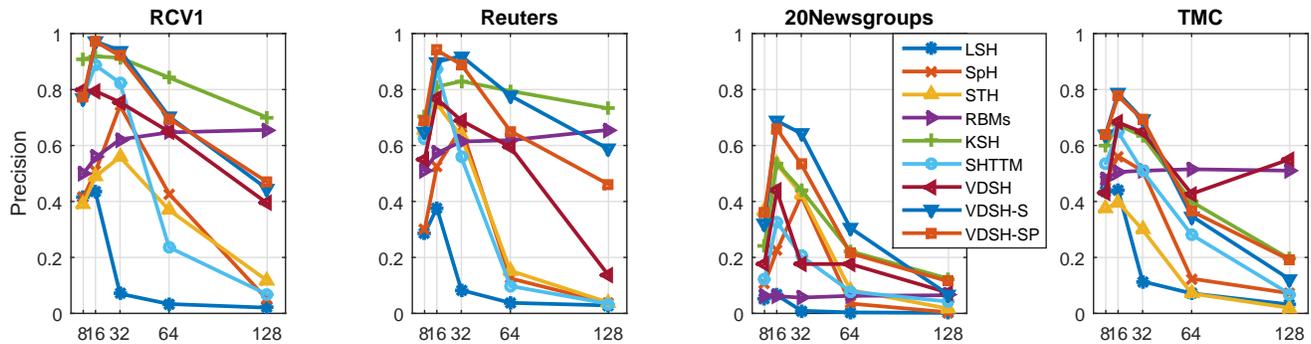}
    \caption{The Precision within the Hamming distance of 2 on four datasets with different hashing bits.} 
\label{fig:hamming_dist_plot}
\end{figure*}

\subsection{Evaluation Metrics}

To evaluate the effectiveness of hashing code in similarity search, each document in the test set is used as a query document to search for similar documents based on the Hamming distance (i.e., number of different bits) between their hashing codes. Following the prior work in text hashing \cite{wang2013semantic}, the performance is measured by Precision, as the ratio of the number of retrieved relevant documents to the number of all retrieved documents. The results are averaged over all the test documents.

There exist various ways to determine whether a retrieved document is relevant to the given query document. In SpH \cite{weiss2009spectral}, the $K$ closest documents in the original feature space are considered as the relevant documents. This metric is not desirable since the similarity in the original feature space may not well reflect the document semantic similarity. Also, it is hard to determine a suitable $K$ for the cutoff threshold. Instead, we adopt the methodology used in KSH \cite{wang2010semi}, SHTTM \cite{wang2013semantic} and other prior work \cite{wang2010semi}, that is a retrieved document that shares any common test label with the query document is regarded as a relevant document. 

\section{Experimental Results}

\subsection{Baseline Comparison}

Table \ref{table:main_experiments} shows the results of different methods over various numbers of bits on the four testbeds. We have several observations from the results. First of all, the best results at different bits are all achieved by VDSH-S or VDSH-SP. They consistently yield better results than all the baselines across all the different numbers of bits. All the improvements over the baselines are statistically significant based on the paired t-test (p-value $< 0.01$). VDSH-S and VDSH-SP produce comparable results between them. Adding private variables does not always help because it increases the model flexibility which may lead to overfitting to the training data. This probably explains why VDSH-SP generally yield better performance when the number of bits is 8 which corresponds to a simpler model. 

Secondly, the supervised hashing techniques (i.e., VDSH-S, VDSH-SP, KSH) outperform the unsupervised methods on the four datasets across all the different bits. These results demonstrate the importance of utilizing supervisory signals for text hashing. However, the unsupervised model, STHs, outperforms SHTTM on the original 20 categories Newsgroups. One possible explanation is that SHTTM depends on LDA to learn an initial representation. But many categories in Newsgroup are correlated, LDA could assign similar topics to documents from related categories (i.e. Christian, Religion). Hence SHTTM may not effectively distinguish two related categories. Evidently, SHTTM and KSH deliver comparable results except on the 20Newsgroups testbed. It is worth noting that there exist substantial gaps between the supervised and unsupervised proposed models (VDSH-S and VDSH-SP vs VDSH) across all the datasets and configurations. The label information seems remarkably useful for guiding the deep generative models to learn effective representations. This is probably due to the high capacity of the neural network component which can learn subtle patterns from supervisory signals when available.

Thirdly, the performance does not always improve when the number of bits increases. This pattern seems quite consistent across all the compared methods and it is likely the result of model overfitting, which suggests that using a long hash code is not always helpful especially when training data is limited. Last but not least, the testbeds may affect the model performance. All the best results are obtained on the RCV1 dataset whose size is much larger than the other testbeds. These results illustrate the importance of using a large amount of data to train text hashing models.

It is worth noting that some of the baseline results are different from what were reported in the prior work. This is due to the data preprocessing. For example, \cite{wang2013semantic} combined some categories in 20Newsgroup to form 6 broader categories in their experiments while we use all the original 20 categories for evaluation. \cite{zhang2010self} focused on single-label documents by discarding the documents appearing in more than one category while we use all the documents in the corpus.

\subsection{Retrieval with Fixed Hamming Distance}
In practice, IR systems may retrieve similar documents in a large corpus within a fixed Hamming distance radius to the query document. In this section, we evaluate the precision for the Hamming radius of 2. Figure \ref{fig:hamming_dist_plot} shows the results on four datasets with different numbers of hashing bits. We can see that the overall best performance among all nine hashing methods on each dataset is achieved by either VDSH-S or VDSH-SP at the 16-bit. In general, the precision of most of the methods decreases when the number of hashing bits increases from 32 to 128. This may be due to the fact that when using longer hashing bits, the Hamming space becomes increasingly sparse and very few documents fall within the Hamming distance of 2, resulting in more queries with precision 0. Similar behavior is also observed in the prior work such as KSH \cite{liu2012supervised} and SHTTM \cite{wang2013semantic}. A notable exception is Stacked RBMs whose performance is quite stable across different numbers of bits while lags behind the best performers. 

\begin{table}[t]
\centering
\begin{tabular}{ c|cc|cc }
  \hline
 & \multicolumn{2}{|c}{RCV1} & \multicolumn{2}{|c}{Reuters}\\
  \hline
 & Median & Sign & Median & Sign \\
  \hline
  VDSH & 0.8481 & 0.8514 & 0.7753 & 0.7851 \\
  VDSH-S & 0.9801 & 0.9804 &  0.9337 & 0.9284 \\
  VDSH-SP & 0.9788 & 0.9794 & 0.9283 & 0.9346 \\
 \hline
 & \multicolumn{2}{|c}{20Newsgroups} & \multicolumn{2}{|c}{TMC}\\
  \hline
 & Median & Sign & Median & Sign \\
 \hline
  VDSH & 0.4354 & 0.4267 & 0.7108 & 0.7162 \\
  VDSH-S & 0.7564 & 0.7563 &  0.7883 & 0.7879 \\
  VDSH-SP & 0.6913 & 0.6574 & 0.7891 & 0.7761 \\
 \hline
\end{tabular}
\caption{Precision$@$100 of using different thresholding functions (Median vs Sign) for the proposed models on four testbeds with the 32-bit hash code}
\label{table:threshold_results}
\end{table}

\subsection{Effect of Thresholding}
\label{thresholding}

Thresholding is an important step in hashing to transform a continuous document representation to a binary code. We investigate two popular thresholding functions: Median and Sign, which are introduced in Section \ref{binhcode}. Table \ref{table:threshold_results} contains the precision results of the proposed models with the 32-bit hash code on the four datasets. As we can see, the two thresholding functions generate quite similar results and their differences are not statistically significant, which indicates all the proposed models, whether being unsupervised or supervised, are not sensitive to the thresholding methods.

\subsection{Effect of Term Weighting Schemes}
\label{effectTFIDF}

\begin{figure}[t]  
\vspace*{-0.5cm}
\hspace*{-0.7cm}
    \includegraphics[scale=0.60]{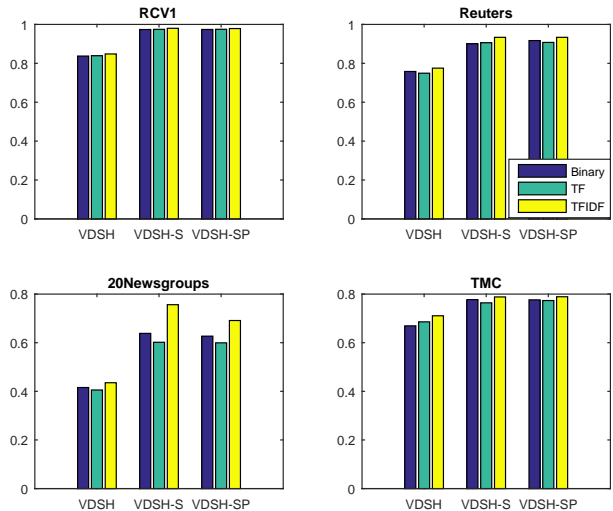}
    \caption{Precision$@$100 for different term weighting schemes on the proposed models with the 32-bit hash code.} 
\label{fig:termweighting}
\end{figure}

In this section, we investigate the effect of term weighting schemes on the performance of the proposed models. Different term weights result in different bag-of-word representations of $\boldsymbol{d}$ as the input to the neural network. Specifically, we experiment with three term weighting representations for documents: Binary, Term Frequency (TF), Term Frequency and Inverse Document Frequency (TFIDF) \cite{manning2008introduction}. Figure \ref{fig:termweighting} illustrates the results of the proposed models with the 32-bit hash code on the four datasets. As we can see, the proposed models generally are not very sensitive to the underlying term weighting schemes. The TFIDF weighting always gives the best performance on all the four datasets. The improvement is more noticeable with VDSH-S and VDSH-SP on 20Newsgroups. The results indicate more sophisticated weighting schemes may capture more information about the original documents and thus lead to better hashing results. One the other hand, all the three models yield quite stable results on RCV1, which suggests that a large-scale dataset may help alleviate the shortcomings of the basic term weighting schemes.

\begin{figure*}[t]
\centering
\begin{subfigure}{.5\textwidth}
  \centering
  \includegraphics[scale=0.6]{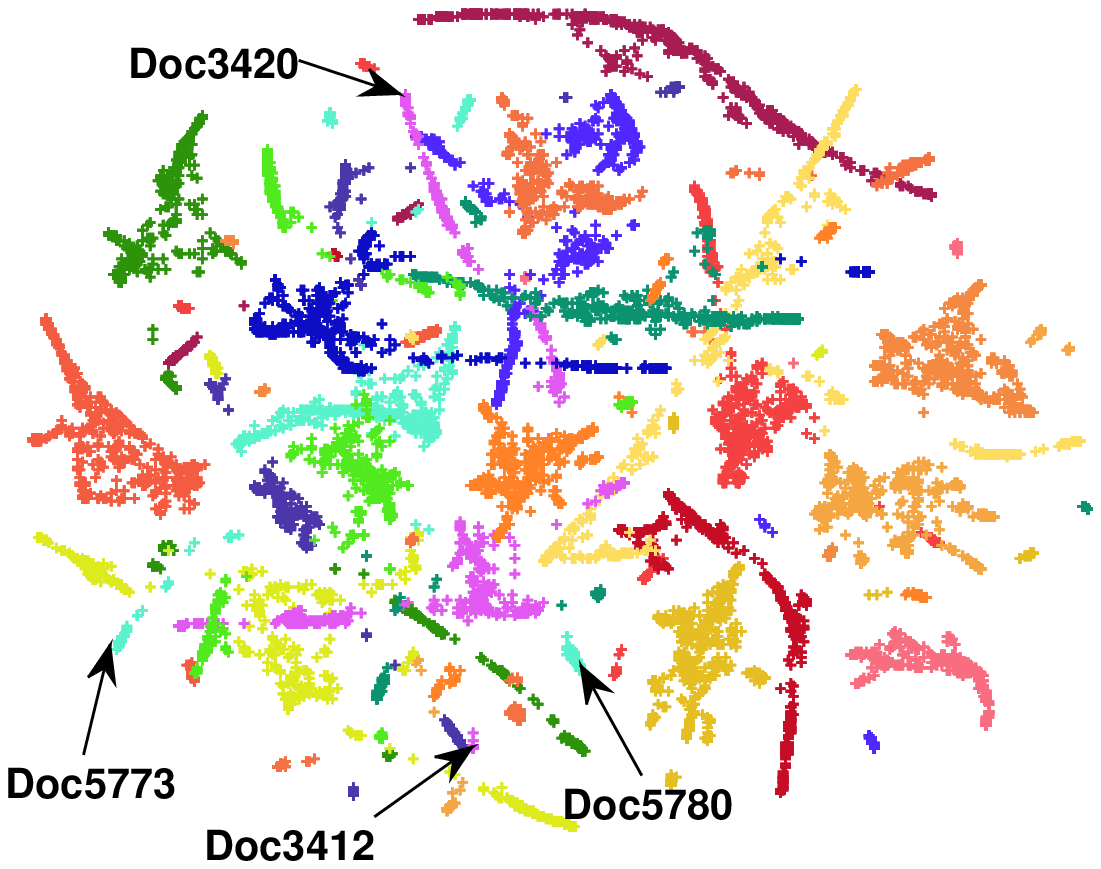}
  \caption{SHTTM}
\end{subfigure}%
\begin{subfigure}{.5\textwidth}
  \centering
  \includegraphics[scale=0.6]{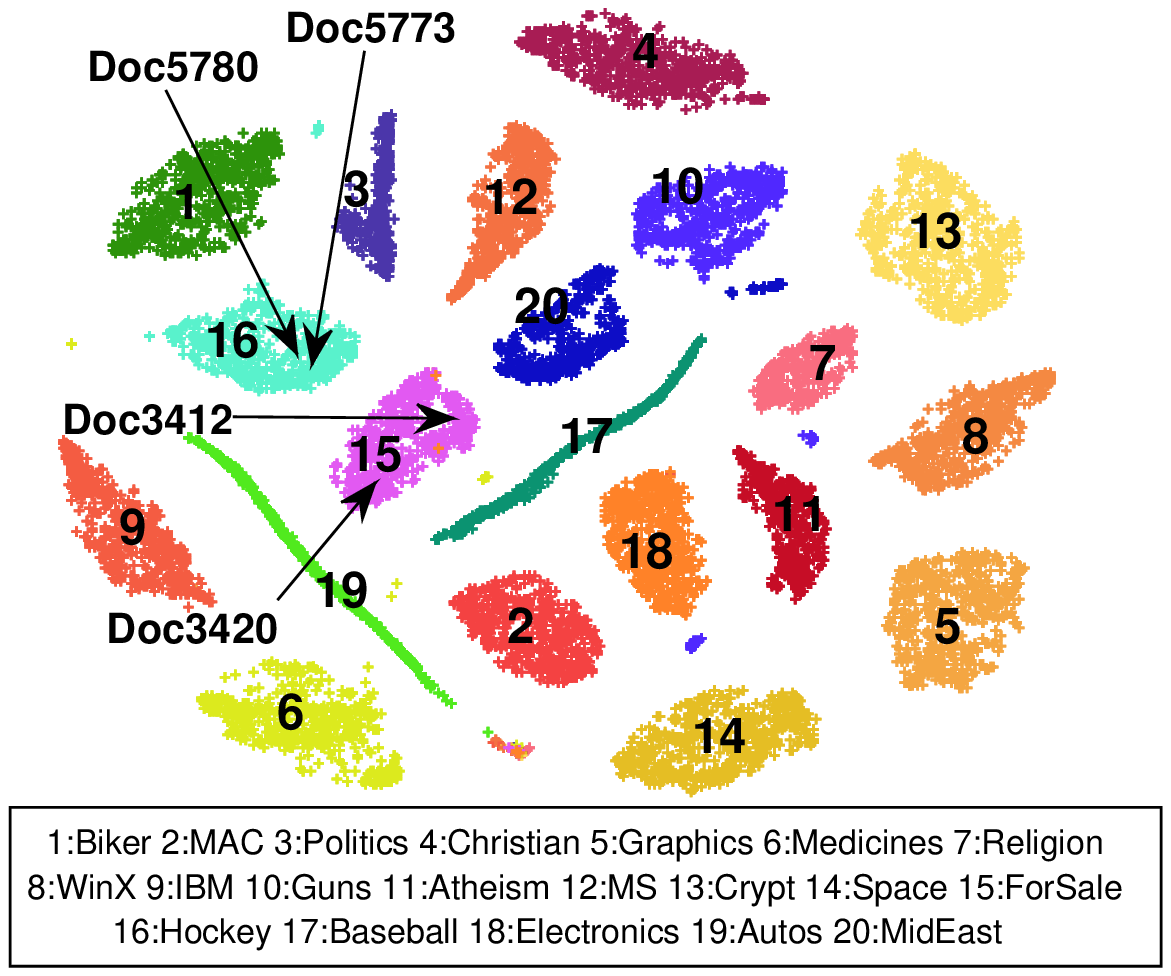}
  \caption{VDSH-S}
  \label{fig:sub2}
\end{subfigure}
\caption{Visualization of the 32-dimensional document latent semantic vectors by SHTTM and VDSH-S on the 20Newsgroup dataset using t-SNE. Each point represents a document and different colors denote different categories based on the ground truth. In (b)VDSH-S, each number is a category ID and the corresponding categories are shown below the plot.}
\label{fig:tsne}
\end{figure*}

\subsection{Qualitative Analysis}
\label{qualitative}

In this section, we visualize the low-dimensional representations of the documents and see whether they can preserve the semantics of the original documents. Specifically, we use t-SNE\footnote{\url{https://lvdmaaten.github.io/tsne/}} \cite{maaten2008visualizing} to generate the scatter plots for the document latent semantic vectors in 32-dimensional space obtained by SHTTM and VDSH-S on the 20Newsgroup dataset. Figure \ref{fig:tsne} shows the results. Here, each data point represents a document which is associated with one of the 20 categories. Different colors represent different categories based on the ground truth.

As we can see in Figure (\ref{fig:tsne})(b), VDSH-S generates well separated clusters with each corresponding to a true category (each number in the plot represents a category ID). On the other hand, the clustering structure from SHTTM shown in Figure (\ref{fig:tsne})(a) is much less evident and recognizable. Some closeby clusters in Figure (\ref{fig:tsne})(b) are also semantically related, e.g., Category 7 (Religion) and Category 11 (Atheism); Category 20 (Middle East) and Category 10 (Guns); Category 8 (WinX) and Category 5 (Graphics). 

We further sampled some documents from the dataset and see where they are represented in the plots. Table \ref{table:sampledoc} contains the DocIDs, categories, and subjects of the sample documents. Doc5780 discusses some trade rumor in NHL and Doc5773 is about NHL team leaders. Both documents belong to the category of Hockey and should be close to each other, which can be clearly observed in Figure (\ref{fig:tsne})(b) by VDSH-S. However, these two documents are projected far away from each other by SHTTM as shown in Figure (\ref{fig:tsne})(a). For another random pair of documents Doc3420 and Doc3412 in the plots, VDSH-S also puts them much closer to each other than SHTTM does. These results demonstrate the great effectiveness of VDSH-S in learning low-dimensional representations for text documents.

\begin{table}[t]
\centering
\begin{tabular}{ |c|c|l| }
 \hline
 \textbf{DocId} &\textbf{Category}& \textbf{Title/Subject} \\
 \hline
  Doc5780 & Hockey&Trade rumor: Montreal/Ottawa/Phillie \\
  Doc5773 & Hockey&NHL team leaders in +/- \\
  Doc3420 & ForSale& Books For Sale [Ann Arbor, MI] \\
  Doc3412 & ForSale&*** NeXTstation 8/105 For Sale *** \\
\hline
\end{tabular}
\caption{The titles of the four sample documents in Figure \ref{fig:tsne}}
\label{table:sampledoc}
\end{table}

\section{Conclusions and Future Work}

Text hashing has become an important component in many large-scale information retrieval systems. It attempts to map documents in a high-dimensional space into a low-dimensional compact representation, while preserving the semantic relationship of the documents as much as possible. Deep learning is a powerful representation learning approach and has demonstrated its effectiveness of learning effective representations in a wide range of applications, but there is very little prior work on utilizing it for text hashing tasks. In this paper, we exploit the recent advances in variational autoencoder and propose a series of deep generative models for text hashing. The models enjoy the advantages of both deep learning and probabilistic generative models. They can learn subtle nonlinear semantic representation in a principled probabilistic framework, especially when supervisory signals are available. The experimental results on four public testbeds demonstrate that the proposed supervised models significantly outperform the competitive baselines.  

This work is an initial step towards a promising research direction. The probabilistic formulation and deep learning architecture provide a flexible framework for model extensions. In future work, we will explore deeper and more sophisticated architectures such as Convolutional Neural Network (CNN), Recurrent Neural Network (RNN) \cite{lecun2015deep} autoregressive neural network (NADE) \cite{larochelle2011neural}


for encoder and decoder. These more sophisticated models may be able to capture the local relations (by CNN) or sequential information (by RNN, NADE, MADE) in text. Moreover, we will utilize the probabilistic generative process to sample and simulate new text, which may facilitate the task of Natural Language Generation \cite{reiter2000building}. Last but not least, we will adapt the proposed models to hash other types of data such as images and videos.

\section*{Appendix}
In this section, we show the derivations of the proposed models. Due to the page limit, we only focus on VDSH-SP, the most sophisticated one among the three models. The other two models can be similarly derived.  

The likelihood of document $d$ and labels $Y$ is:

\begin{align*}
&\log P(\boldsymbol{d}, \boldsymbol{Y})=\log \int_{\boldsymbol{s, v}} P(\boldsymbol{d}, \boldsymbol{Y}, \boldsymbol{s}, \boldsymbol{v}) d\boldsymbol{s}d\boldsymbol{v} \\
&=\log \int_{\boldsymbol{s, v}} Q(\boldsymbol{s}, \boldsymbol{v}|\boldsymbol{d}, \boldsymbol{Y}) \frac{P(\boldsymbol{d}, \boldsymbol{Y}, \boldsymbol{s}, \boldsymbol{v})}{Q(\boldsymbol{s}, \boldsymbol{v}|\boldsymbol{d}, \boldsymbol{Y})} d\boldsymbol{s}d\boldsymbol{v} \\
\end{align*}
Based on the Jensen's Inequality \cite{goodfellow2016deep},
\begin{eqnarray}
&&\log P(\boldsymbol{d}, \boldsymbol{Y}) \ge \E_{Q(\boldsymbol{s,v})}[\log P(\boldsymbol{d}, \boldsymbol{Y}, \boldsymbol{s}, \boldsymbol{v}) - \log Q(\boldsymbol{s}, \boldsymbol{v}|\boldsymbol{d}, \boldsymbol{Y})] \nonumber\\
&=& \E_{Q(\boldsymbol{s,v})}[\log P(\boldsymbol{d}|\bm{s,v})P(\boldsymbol{Y}|\bm{s})] + \E_{Q(\boldsymbol{s,v})}[\log P(\boldsymbol{s}) \nonumber \\
&+& \log P(\boldsymbol{v}) - \log Q(\boldsymbol{s,v}|\boldsymbol{d,Y})] \label{line1} \\
&=& \E_{Q(\boldsymbol{s,v})}[\log P(\boldsymbol{d}|\boldsymbol{s,v})P(\boldsymbol{Y}|\boldsymbol{s})] 
+ \E_{Q(\boldsymbol{s,v})}[\log P(\boldsymbol{s}) - \log Q(\boldsymbol{s}|\boldsymbol{d})] \nonumber \\
&+& \E_{Q(\boldsymbol{s,v})}[\log P(\boldsymbol{v}) - \log Q(\boldsymbol{v}|\boldsymbol{d})] \label{line2} \\
&=& \E_{Q(\boldsymbol{s,v})}[\log P(\boldsymbol{d}|\boldsymbol{s,v})P(\boldsymbol{Y}|\boldsymbol{s})] -\infdiv{Q(\boldsymbol{s}|\boldsymbol{d})}{P(\boldsymbol{s})} \nonumber \\ &-&\infdiv{Q(\boldsymbol{v}|\boldsymbol{d})}{P(\boldsymbol{v})} \label{line3} \\
&=& \E_{Q(\boldsymbol{s,v})}[\log P(\boldsymbol{d}|\boldsymbol{s,v})] + \E_{Q(\boldsymbol{s,v})}[\log P(\boldsymbol{Y}|\boldsymbol{s})] \nonumber \\
&-&\infdiv{Q(\boldsymbol{s}|\boldsymbol{d})}{P(\boldsymbol{s})} -\infdiv{Q(\boldsymbol{v}|\boldsymbol{d})}{P(\boldsymbol{v})} \label{line4}\\
&=&\E_{Q(\boldsymbol{s,v})} \big[ \sum_{i=1}^N \log P(\boldsymbol{w}_i | f(\boldsymbol{s}+\boldsymbol{v}; \theta))+\sum_{j=1}^L \log P(\boldsymbol{y}_j | f(\boldsymbol{s}; \tau))\big] \nonumber\\
&-& \infdiv{Q(\boldsymbol{s}|\boldsymbol{d}; \phi)}{P(\boldsymbol{s})}-\infdiv{Q(\boldsymbol{v}|\boldsymbol{d}; \phi)}{P(\boldsymbol{v})} \label{line5}
\end{eqnarray}
In Eqn.(\ref{line1}), we factorize the joint probability based on the generative process. Thus, $P(\boldsymbol{d,Y,s,v}) = P(\boldsymbol{d}|\boldsymbol{s,v})P(\boldsymbol{Y}|\boldsymbol{s})P(\bm{s})P(\boldsymbol{v})$. In Eqn.(\ref{line2}), the variational distribution, $Q(\boldsymbol{s,v}|\boldsymbol{d,Y})$ is equal to the product of $Q(\boldsymbol{s}|\boldsymbol{d})$ and $Q(\boldsymbol{v}|\boldsymbol{d})$ by assuming the conditional independence of$ \boldsymbol{s, v, Y}$ given $\boldsymbol{d}$. Eqn.(\ref{line3}) and Eqn.(\ref{line4}) are the results of rearranging and simplifying terms in Eqn.(\ref{line2}). Plugging the individual words and labels, we obtain the final lowerbound objective function in Eqn.(\ref{line5}) (also in Eqn.(\ref{lowerbound3})).

Because of the Gaussian assumptions on latent semantic vector $\boldsymbol{s}$ and latent private variable $\boldsymbol{v}$, the two KL divergences in Eqn.(\ref{line5}) have analytic forms. We let $\boldsymbol{\mu}_s$ and $\boldsymbol{\sigma}_s$ are mean and standard deviation of $\boldsymbol{s}$. $\boldsymbol{\mu}_v$ and $\boldsymbol{\sigma}_v$ are similar defined. We use subscript $k$ to denote the $k^{th}$ element of the vector. The following derivation is an analytical form for a single KL divergence term:
\begin{align} \label{eq:kl_terms}
&\infdiv{Q(\boldsymbol{s}|\boldsymbol{d}; \phi)}{P(\boldsymbol{s})} \nonumber
= \E_{Q(\boldsymbol{s})}[\log P(\boldsymbol{s})] - \E_{Q(\boldsymbol{s})}[\log Q(\boldsymbol{s}|\boldsymbol{d}; \phi)] \nonumber \\
&= \frac{1}{2}\sum_{k=1}^K(1+\log \boldsymbol{\sigma}_{s,k}^2-\boldsymbol{\mu}_{s,k}^2-\boldsymbol{\sigma}_{s,k}^2)
\end{align}

$\infdiv{Q(\boldsymbol{v}|\boldsymbol{d}; \phi)}{P(\boldsymbol{v})}$ can be derived in the same way. The expectation terms in Eqn.(\ref{line5}) do not have a closed form solution, but we can approximate them by the Monte Carlo simulation as follows:

\begin{align} \label{eq:expected_terms}
&\E_{Q(\boldsymbol{s,v})}[\log P(\boldsymbol{d}|f(\bm{s}+\bm{v}; \bm{\theta}))] + \E_{Q(\boldsymbol{s,v})}[\log P(\boldsymbol{Y}|\boldsymbol{s})] \nonumber \\
&= \frac{1}{M}\sum_{m=1}^M \bigg( \log P(\boldsymbol{d}|f(\bm{s}^{(m)}+\bm{v}^{(m)}; \bm{\theta})) + \log P(\boldsymbol{Y}|\boldsymbol{s}^{(m)}) \bigg)
\end{align}

The superscript $m$ denotes the $m^{th}$ sample. By shift and scale transformation, we have $\boldsymbol{s}^{(m)} = \boldsymbol{\epsilon}_s^{(m)} \odot \boldsymbol{\sigma}_s + \boldsymbol{\mu}_s$. We denote $\boldsymbol{\epsilon}_s$ as a sample drawn from a standard multivariate normal and $\odot$ is an element-wise multiplication. Also, $\boldsymbol{v}^{(m)}$ is obtained in the same way, $\boldsymbol{v}^{(m)} = \boldsymbol{\epsilon}_v^{(m)} \odot \boldsymbol{\sigma}_v + \boldsymbol{\mu}_v$. By using this trick, we can obtain multiple samples of $\boldsymbol{\epsilon}$ and feed them as the deterministic input to the neural network. The model becomes an end-to-end deterministic deep neural network with the following objective function:

\begin{align} \label{eq:final_obj_func}
& \mathcal{L} = \frac{1}{M}\sum_{m=1}^M \bigg( \log P(\boldsymbol{d}|f(\bm{s}^{(m)}+\bm{v}^{(m)}; \bm{\theta})) + \log P(\boldsymbol{Y}|\boldsymbol{s}^{(m)})\bigg) \nonumber \\
&+ \frac{1}{2}\sum_{k=1}^K(2+\log \boldsymbol{\sigma}_{s,k}^2-\boldsymbol{\mu}_{s,k}^2-\boldsymbol{\sigma}_{s,k}^2 + \log \boldsymbol{\sigma}_{v,k}^2-\boldsymbol{\mu}_{v,k}^2-\boldsymbol{\sigma}_{v,k}^2)
\end{align}





\begin{acks}
The authors would like to thank Mengwen Liu, for her many helpful comments and suggestions on the first draft, Travis Ebesu for his drawing of the model architectures in Figure \ref{fig:arch}, and anonymous reviewers for valuable comments and feedback.
\end{acks}

\bibliographystyle{abbrv}
\bibliography{sigproc}  

\end{document}